\documentclass[conference]{IEEEtran}
\IEEEoverridecommandlockouts
\usepackage{cite}
\usepackage{amsmath,amssymb,amsfonts}
\usepackage{algorithmic}
\usepackage{graphicx}
\usepackage{array}
\usepackage{textcomp}
\usepackage{xcolor}
\def\BibTeX{{\rm B\kern-.05em{\sc i\kern-.025em b}\kern-.08em
    T\kern-.1667em\lower.7ex\hbox{E}\kern-.125emX}}
\begin{document}

\title{Power Domain Sparse Dimensional Constellation Multiple Access (PD-SDCMA): A Novel PD-NOMA for More Access Users}

\author{
	\IEEEauthorblockN{
		Zihan Li\IEEEauthorrefmark{1}\IEEEauthorrefmark{2}, 
		Youzhi Li\IEEEauthorrefmark{1}\IEEEauthorrefmark{2}\IEEEauthorrefmark{3}, 
		Chenyu Liu\IEEEauthorrefmark{1}\IEEEauthorrefmark{2}\IEEEauthorrefmark{3} 
		and Yuhao Lian\IEEEauthorrefmark{1}\IEEEauthorrefmark{4}} 
	\IEEEauthorblockA{\IEEEauthorrefmark{2}Chongqing University of Posts and Telecommunications, Chongqing 400065, China \\ \IEEEauthorrefmark{3}Department of Electronic and Electrical Engineering, Brunel University London, Uxbridge, Middlesex UB8 3PH, UK\\ 
    \IEEEauthorrefmark{4}College of Engineering, Zhejiang University, Hangzhou 310027, China\\
    Email: 2279372@brunel.ac.uk, 2022210299@stu.cqupt.edu.cn, 2378898@brunel.ac.uk, lianyuhao@ieee.org}
	\IEEEauthorblockA{\IEEEauthorrefmark{1}These authors contributed  equally to this work}
}

\maketitle

\begin{abstract}
With the advent of the 6G mobile communication network era, the existing non-orthogonal multiple-access (NOMA) technology faces the challenge of high successive interference in multi-user scenarios, which limits its ability to support more user access. To address this, this paper proposes a novel power-domain sparse-dimensional constellation multiple-access scheme (PD-SDCMA). Through the signal space dimension selection strategy (S2D-strategy), this scheme sparsely superposes low-dimensional constellations onto high-dimensional signal spaces, and reduces the high-order interference caused by SC by taking advantage of the non-correlation between dimensions. Specifically, PD-SDCMA reduces the successive interference between users by sparsifying the dimension allocation of constellation superposition and designs a sparse superposition method based on the theory of vector space signal representation. Simulation results show that, under the AWGN channel, PD-SDCMA significantly outperforms the traditional PD-NOMA in terms of the number of supported users under QPSK and 16QAM modulations, and also has better BER performance. This paper provides a new solution for efficient spectrum utilization in future scenarios with large-scale user access.
\end{abstract}


\section{Introduction}
In the current 5th generation mobile network (5G) era, these orthogonal schemes such as Frequency Division Multiple Access \cite{b1}, Time Division Multiple Access (TDMA) \cite{b2}, Code Division Multiple Access (CDMA)\cite{b3} and Orthogonal Frequency Division Multiple Access (OFDMA)\cite{b4} are increasingly inadequate for meeting the escalating requirements of modern communication systems, particularly as the spectrum resource scarcity becomes more pronounced. In this context, Power-domain Non-Orthogonal Multiple Access (PD-NOMA) has garnered significant attention as a promising candidate for future-generation mobile networks\cite{b5}. Unlike orthogonal multiple access methods, PD-NOMA allows multiple users to share the same time-frequency resource block by superimposing their signals in the power domain. This is achieved through advanced techniques such as Successive Interference Cancellation (SIC)\cite{b6}, where users decode and subtract the signals of other users before decoding their own messages. Meanwhile, Code-domain Non-Orthogonal Multiple Access (CD-NOMA) like Sparse Code Multiple Access (SCMA)\cite{b7}, Pattern Division Multiple Access (PDMA)\cite{b8}, and Multi-user Shared Access (MUSA)\cite{b9} is based on the codebook design approach and employs the message passing algorithm (MPA) to separate signals from different users, enabling large-scale device connections. These approaches enable a more efficient utilization of the available spectrum, thereby enhancing spectral efficiency and accommodating a larger number of users within the same bandwidth. 

However, as the 6th generation mobile network (6G) era dawns\cite{b10}, the anticipated exponential increase in data traffic, coupled with the diverse and dynamic requirements of various applications—including enhanced mobile broadband plus (eMBB-Plus)\cite{b11}, secure ultra-reliable low latency communication (SURLLC)\cite{b12}, three dimensional integrated communication (3D-InteCom)\cite{b13}, unconventional data communication (UCDC)\cite{b15}, and big communication (BigCom)\cite{b15}—poses significant challenges to above multiple access techniques. Specifically, CD-NOMA requires consideration of more factors in multi-user scenarios due to the complexity of codebook design. In PD-NOMA, as the number of users increases, the superposition leads to an increase in successive interference, making it difficult to demodulate the correct information. Ultimately, NOMA is limited in multi-user scenarios, particularly for PD-NOMA. 

To accommodate such demands, PD-NOMA is integrated with various methods to adapt to more flexible scenarios as follows. \cite{b16} combines MIMO with PD-NOMA to further enhance spectral efficiency and reduces the inter-cell interference (ICI) problem in multi-cell scenarios. \cite{b17} combines reconfigurable intelligent surfaces (RIS) with it and applies it to satellite-aerial-terrestrial networks (SATN), improving coverage, connectivity, and reliability. \cite{b18} empowers PD-NOMA with semantic communication, enhancing resource efficiency and addressing the "near-far user rate disparity" problem. However, the above methods do not address the fundamental issue of high serial interference caused by the increase in the number of users, which poses challenges to the future implementation of PD-NOMA in scenarios with more users.

To reduce the high serial interference in PD-NOMA, we have proposed a novel PD-NOMA scheme for more access users called power domain sparse dimensional constellation multiple access (PD-SDCMA). This technology adopts a dimension selection strategy based on high-dimensional space superposition: that is, low-dimensional constellations achieve the superposition of sparse constellation points in a high-dimensional signal space. By leveraging the sparse uncorrelated dimensions, it reduces the high-order interference caused by superposition coding (SC). The main contributions of this paper are summarized as follows:

\begin{itemize}
\item We first proposed the PD-SDCMA scheme for supporting more access users, where sparsification is performed in the dimension of constellation superposition to reduce the interference caused by SC.
\item We further elaborate on the fundamental theory of the signal space dimension selection strategy (S2D-strategy) and sparse superposition method for PD-SDCMA from the perspective of the vector space representation of signals.
\item Moreover, we conducted simulations in the Additive White Gaussian Noise (AWGN) channel with different numbers of users, based on two-dimensional constellation (QPSK,16QAM) modulation schemes. These simulations verified the advantage of the proposed PD-SDCMA scheme in increasing the number of access users.
\end{itemize}

This paper is organized as follows: In Section \ref{S2}, we present a comprehensive description of the PD-SDCMA system model, where particular emphasis is given to theory of the signal space dimension selection strategy (S2D-strategy) and sparse superposition method for PD-SDCMA from the perspective of the vector space representation of signals. Then, in Section \ref{S3}, we perform simulations to compare the supporting access users with PD-NOMA in an AWGN channel. Finally, Section \ref{S4} concludes the paper, summarizing the results obtained from the simulations and the main contributions of the article.

\section{PD-SDCMA OPERATING PRINCIPLE}\label{S2}

This Section introduces the prerequisite knowledge—Vector space representation of signals required for the proposed PD-SDCMA, along with the description and analysis of theory of the S2D-strategy and sparse superposition method for PD-SDCMA.

\subsection{Vector space representation of signals}\label{S21}

\begin{figure}[htbp]
\centering % 表示居中
\includegraphics[height=10cm,width=6.923cm]{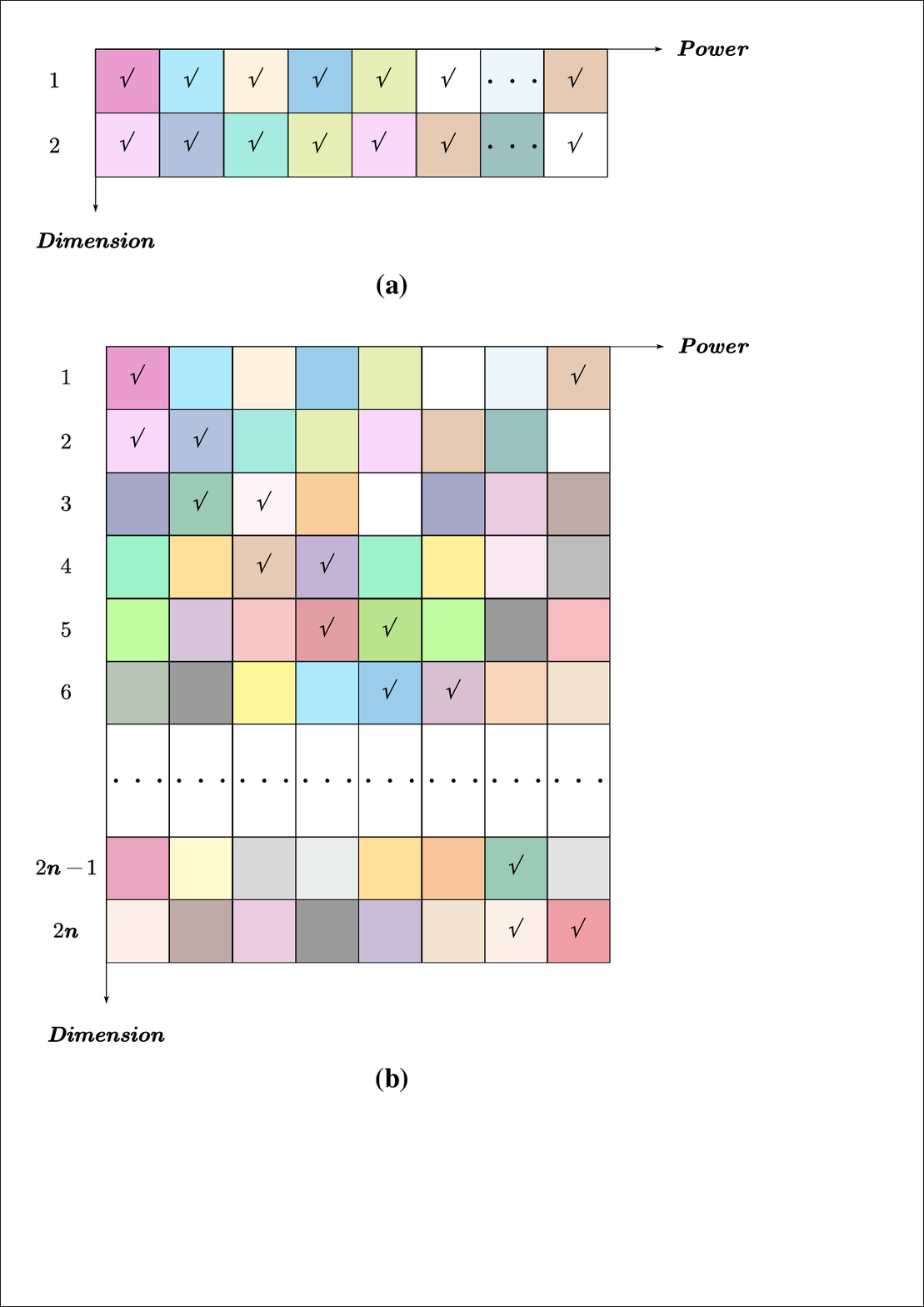}
\caption{Schematic illustration of S2D-strategy.}
\label{fig1}
\end{figure}

\begin{figure}
\centering % 表示居中
\includegraphics[height=4.5cm,width=8cm]{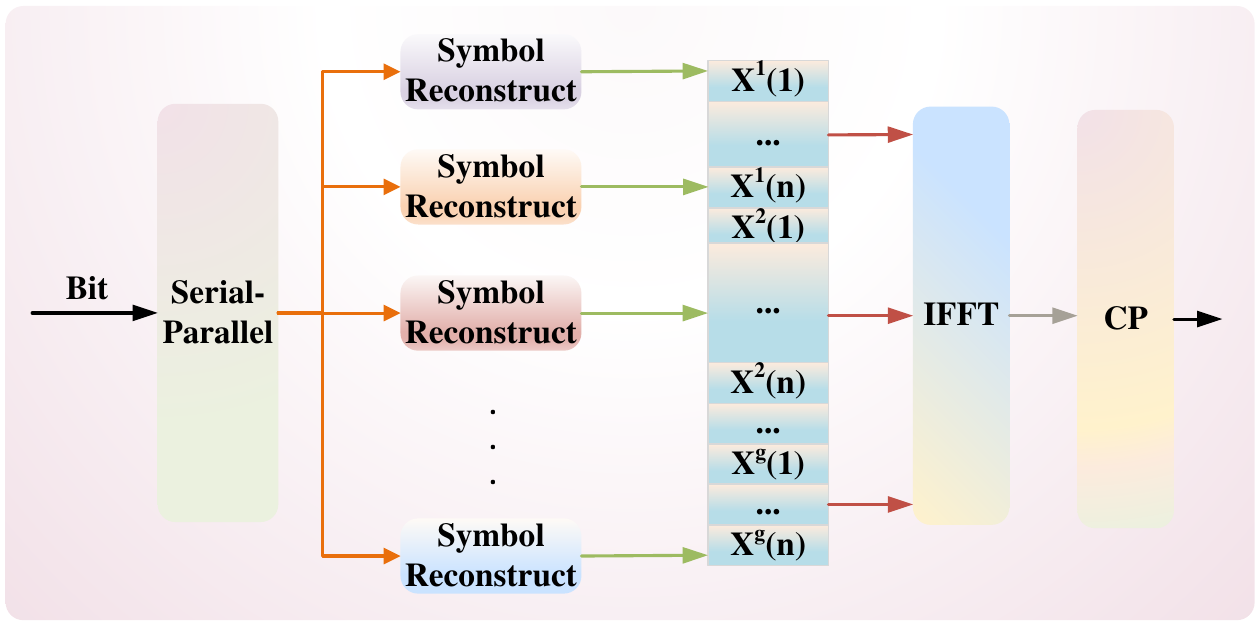}
\caption{The DSP processing procedure at the transmitter.}
\label{fig2}
\end{figure}

\begin{figure*}
\centering % 表示居中
\includegraphics[height=4.2cm,width=18cm]{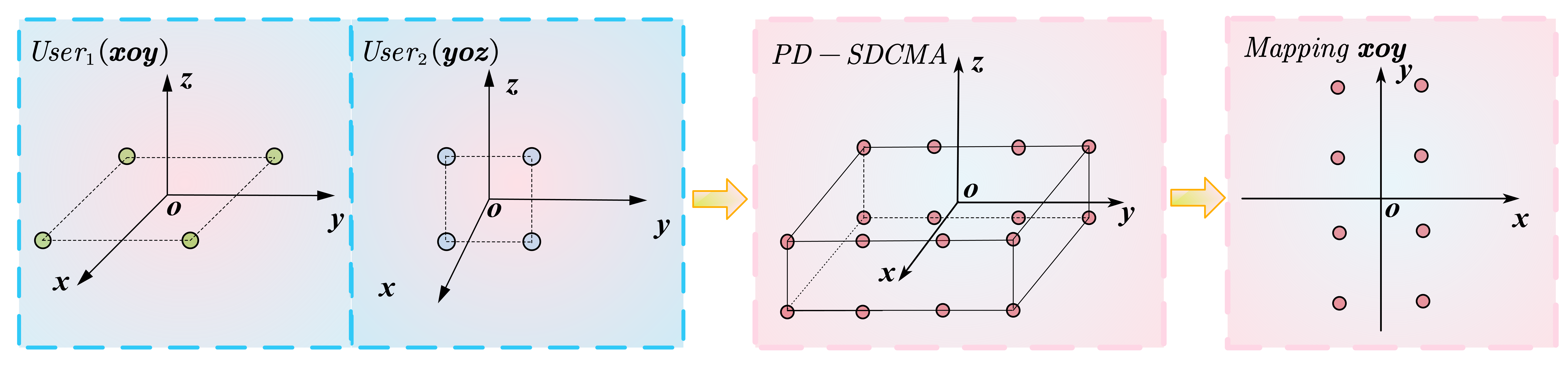}
\caption{Schematic illustration of S2D-strategy.}
\label{fig3}
\end{figure*}

In the vector space analysis theory of digital signals, a point in the vector space represents a signal waveform. The orthonormal basis functions in the vector space correspond to the respective spatial dimensions. If $N$ functions $f_k(t)$, where $k=1,2,...,N$, satisfy:

\begin{equation}
\int_{-\infty}^{\infty} f_{n}(t)f_{m}(t)=
\begin{cases}
0, & m\neq n \\
1, & m = n
\end{cases}
\end{equation}
 The functions are orthogonal to each other, and the function set $\{f_{k}(t), k = 1,2,\cdots,N\}$ is called an orthonormal basis function set. Suppose a signal waveform $S(t)$ can be represented as a linear combination of $f_k(t)$, that is:

\begin{equation}
S(t)=\sum_{k = 1}^{N}s_{k}f_{k}(t)
\end{equation}
The position of a signal in the vector space can be determined by the projections of the signal waveform $S(t)$ onto $\{f_{k}(t), k = 1,2,\cdots,N\}$. Therefore, the signal waveform $S(t)$ can be represented in an $N$-dimensional vector space as:

\begin{equation}
\boldsymbol{S}=[s_{1},s_{2},\cdots,s_{N}]
\end{equation}

In signal analysis, the above-mentioned  $N$-dimensional vector space is called an $N$-dimensional signal space. The $M$ points in the  $N$-dimensional signal space onto which the signal waveform  is mapped are called the constellation of an  $M$-ary signal. In order to represent a signal space with more than two dimensions, the linear combination of $S(t)$ must contain more than two orthonormal basis functions. When a signal waveform is mapped into a signal space, the in-phase and quadrature components of a carrier at the same frequency can only be represented as two orthonormal basis functions, that is, two dimensions in the signal space. Therefore, a signal space with more than two dimensions is composed of the in-phase and quadrature components of multiple orthogonal carriers. The orthogonal carriers must satisfy:

\begin{equation}
\int_{0}^{T_{s}}\cos(2\pi f_{i}t+\varphi_{i})\cos(2\pi f_{j}t+\varphi_{j})dt = 0
\end{equation}
where $T_s$ is the symbol-period time, $f_i$ and $\varphi_i$ are the frequency and phase of the $i_{th}$ carrier, respectively, and $f_j$ and $\varphi_j$ are the frequency and phase of the $j_{th}$  carrier, respectively. Based on this, if the traditional two-dimensional orthogonal constellation is mapped into an  $N$-dimensional signal space, the signal waveform $S(t)$ can be expressed as:

\begin{equation}
\boldsymbol{S}=[s_{1},s_{2},0,0,\cdots,0]
\end{equation}
where $\boldsymbol{S}$ contains $N-2$ zeros, corresponding to the in-phase and quadrature components of the orthogonal carriers in that dimension. Therefore, the traditional two-dimensional QPSK constellation can be represented in a four-dimensional signal space as:

\begin{equation}
\boldsymbol{S}=[s_{1},s_{2},0,0], s_{1}, s_{2} \in \{ \pm 1\}
\end{equation}

\begin{figure*}
\centering % 表示居中
\includegraphics[height=5.5cm,width=18cm]{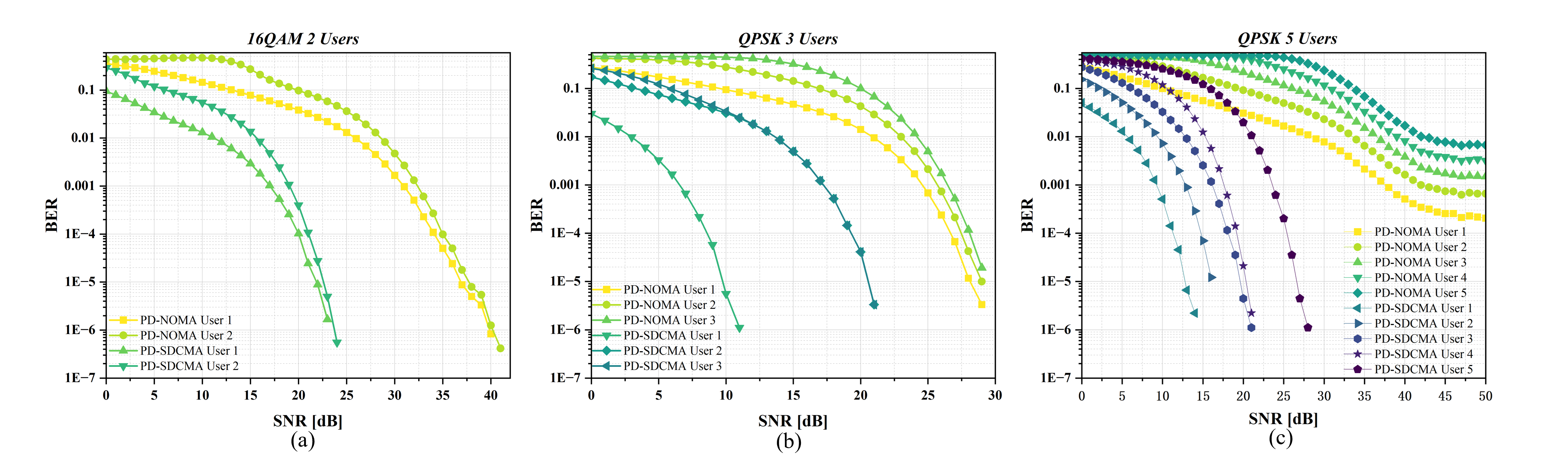}
\caption{The BER performance of different multiple-access schemes as a function of the SNR.}
\label{fig4}
\end{figure*}

\subsection{S2D-strategy for sparse superposition}
Therefore, based on the above-mentioned analysis, all constellations can be mapped to different dimensions of the signal space, achieving sparse superposition from the perspective of constellation dimensions. This constellation-dimension mapping approach is referred to as the signal space dimension selection strategy (S2D-strategy). 

As illustrated in Fig.\ref{fig1}(a), the S2D-strategy of PD-NOMA is depicted, where the horizontal axis represents power, and the vertical axis represents dimension. PD-NOMA employs a 2D constellation for full-dimensional superposition at different power levels, corresponding to the superposition of all resource blocks selected at different power levels. In comparison to Fig.\ref{fig1}(b), PD-SDCMA positions a lower-dimensional constellation in a higher-dimensional space, utilizing a designated S2D-strategy for sparse power superposition, and the S2D-strategy can be represented by a matrix $S_{g\times q}$:

\begin{equation}
S_{g\times q}=\begin{pmatrix}
1&2&\cdots&q - 1&q\\
2&3&\cdots&q&q + 1\\
\cdots&\cdots&\cdots&\cdots&\cdots\\
g - 1&g&\cdots&q - 3&q - 2\\
g&1&\cdots&q - 2&q - 1
\end{pmatrix}
\end{equation}
matrix $S_{g\times q}$ consists of $g$ rows and $q$ columns, where $g$ rows represent the number of access users, and  $q$ columns represent the use of a constellation with $q$ dimensions. The data in the $i_{th}$ row indicates the dimensions selected for the $i_{th}$ access user. Certainly, given the countless possibilities of S2D-strategy, there are also numerous compositions for the $S$ matrix. 

This paper primarily focuses on the higher-dimensional superposition of 2D constellation points. The strategy involves each user access group superposing only one dimension instead of full-dimensional superposition. In this case, the  $S_{g\times q}$ matrix is specifically configured as a  $S_{g\times 2}$ matrix :

\begin{equation}
S_{g\times2}=\begin{pmatrix}
1&2\\
2&3\\
\cdots&\cdots\\
g - 1&g\\
g&1
\end{pmatrix}
\end{equation}

Based on the analysis in Section \ref{S21} above, each constellation dimension corresponds to the in - phase and quadrature components of a different-frequency carrier. Therefore, the single-user data of each PD-SDCMA is carried in the form of orthogonal carriers, that is, the quadrature/in-phase components of the orthogonal carriers corresponding to the mapping dimensions carry data, while the corresponding components of the remaining orthogonal carriers do not carry data. Therefore, PD-SDCMA waveforms can be generated in the same form as OFDM waveforms. The DSP processing flow at the transmitting end is shown in Fig.\ref{fig2}. First, the high-speed serial bit-stream is converted into multiple sets of parallel data through a Serial-Parallel Converter (S/P conversion). Subsequently, these multiple sets of parallel data undergo constellation mapping to become symbols. According to the S2D-strategy, the symbols are mapped to the corresponding signal space dimensions, and the real/imaginary parts corresponding to the unmapped subspace dimensions are left vacant, which is the symbol reconstruction process. After completing symbol reconstruction, an IFFT operation is performed on the reconstructed symbols.

\begin{equation}
Y = IFFT\{X\} \rightarrow Y(n)=\frac{1}{N}\sum_{k = 1}^{N}X(k)W_N^{-nk}
\end{equation}

A Cyclic Prefix (CP) is inserted into the generated time-domain signal, and then a Parallel-Serial (P-S) conversion is carried out.

\begin{equation}
Y'=[y_0',y_1',y_2',\cdots,y_{N - 2}',y_{N - 1}',y_{CP0}',y_{CP1}',\cdots]
\end{equation}

Finally, the $g$ signals generated by the $g$ access users are summed in the power-domain to generate the transmitted signal.

\begin{align}
Z(t)=\sqrt{P_1}Y_1'(t)&+\sqrt{P_2}Y_2'(t)+\cdots+\sqrt{P_g}Y_g'(t)\\
P_1 + P_2 &+ \cdots+P_g= 1
\end{align}

At the receiving end, Successive Interference Cancellation (SIC) can be employed for successive signal separation and demodulation. Unlike the SIC demodulation in PD-NOMA, the demodulation of PD-SDCMA requires mapping the superimposed joint constellation onto the constellation plane of the accessing user. As shown in Fig.\ref{fig3}, take the sparse superposition of the QPSK constellations of two access groups as an example. The transmitting end assigns high power to the first-access user and low power to the second-access user. In a three-dimensional space, the four constellation points of the first-access user are located on the xoy plane, while the constellation of the second-access user is located on the yoz plane. The joint constellation formed by the superposition of the constellations of the two access groups appears as a cuboid constellation with 16 constellation points in the signal space. When the receiving end performs demodulation, it treats the low-power access group as noise to demodulate the high-power access group, that is, maps the 16 constellation points onto the xoy plane. Since the constellation of the first-access user is not mapped onto the z-axis, the constellation points with the same xy-axis coordinates will coincide when mapped onto the xoy plane. Therefore, the joint constellation, when mapped onto the xoy plane, will be transformed into a constellation similar to 8QAM. The serial interference on the z-axis will not affect the demodulation of the first-access user, and the influence of the first-access user on the demodulation of the second-access user is also smaller.

\section{SIMULATION AND RESULTS ANALYSIS}\label{S3}
In this section, we conducted simulations on (AWGN channels with different numbers of users under given parameters and analyzed the simulation results.

\subsection{Experimental Settings}\label{AA}

\begin{table}[h]
		\caption{Parameter Settings}
		\centering
		\begin{tabular}{p{3.5cm}<{\centering} p{3.5cm}<{\centering}}
		\hline
            \hline
			\textbf{Parameter} & \textbf{Value} \\
            \hline
            Constellation & QPSK \& 16QAM \\

            Number of orthogonal carriers & 256\\

            IFFT/FFT points & 512\\

            Number of symbols & 1000\\

            CP & 0.125\\
		\hline
            \hline
		\end{tabular}
		\label{tab1}  % 表格标签，用于引用
	\end{table}

\begin{table}[htbp]
    \centering
    \caption{S2D-strategy}
    \begin{tabular}{cccc}
        \hline
        \hline
        \textbf{Scenarios}  & \textbf{Constellation} & \textbf{S2D-strategy} & \textbf{Power distribution} \\
        \hline
        2 users & 16QAM & [1 2;2 3] & 16:1\\
        3 users & QPSK & [1 2;2 3;3 1] & 16:4:1\\
        5 users & QPSK & [1 2;2 3;3 4;4 5;5 1] & 256:64:16:4:1\\
        \hline
        \hline
    \end{tabular}
    \label{tab2}  % 表格标签，用于引用
\end{table}

In terms of the experimental parameter settings, the simulation base parameters are presented in Table \ref{tab1}. The simulation mainly focuses on two constellation mappings, namely QPSK and 16QAM. The S2D-strategies corresponding to these two constellations scenarios and the appropriate power-distribution ratios are presented in Table \ref{tab2}. Based on the above experimental setup, the simulation verifies the effectiveness of the proposed PD-SDCMA in increasing the number of access users and the capacity of the communication system.

\subsection{RESULTS ANALYSIS}\label{BB}
To evaluate the effectiveness of the proposed PD-SDCMA scheme, The Fig.\ref{fig4} illustrates the bit-error-rate (BER) performance of different multiple-access schemes as a function of the signal-to-noise ratio (SNR). Fig.\ref{fig4} (a) shows the BER performance for 16QAM with two users, comparing PD-NOMA and PD-SDCMA. Fig.\ref{fig4} (b) and (c) present the BER results for QPSK with three and five users, respectively. Specifically, Figure 4(a) indicate that, under BER=$1\times10^{-3}$, the SNR corresponding to user 1 and user 2 in PD-SDCMA are approximately 17dB and 19dB respectively. In contrast, the SNR of user 1 and user 2 in PD-NOMA are approximately 31dB and 32.5dB respectively. This significantly enhances the anti-interference ability of each user. This advantage stems from the fact that PD-SDCMA distributes user signals in orthogonal high-dimensional spaces through the sparse dimension superposition strategy, effectively reducing the mutual interference caused by SC and thus improving the accuracy of signal separation.

Fig.\ref{fig4}(b) and (c) further verify the robustness of PD-SDCMA in multi-user scenarios. Under QPSK modulation, when the number of users increases to three, compared with PD-NOMA, when $1\times10^{-3}$, the SNR gains corresponding to user 1, user 2, and user 3 in PD-SDCMA are approximately 17 dB, 7.7 dB, and 8.7 dB, respectively. In the five-users scenario, the increase in the number of access users leads to a reduction in the constellation minimum Euclidean distance (MED). As a result, even with QPSK constellation mapping, PD-NOMA cannot ensure that BER performance of each access user meets the requirements of the reliable transmission threshold. In contrast, PD-SDCMA can still guarantee reliable transmission for five access users. This result confirms the effectiveness of the sparse dimension allocation strategy in suppressing successive interference-by sparsely mapping user constellations to non-correlated dimensions in high-dimensional spaces, the mutual coupling during signal decoding is reduced, thus significantly improving the multi-user capacity.

\section{Conclusion}\label{S4}

Aiming at the problem of severe successive interference of traditional PD-NOMA in multi-user scenarios, this paper proposes a PD-SDCMA scheme based on sparse superposition in high-dimensional signal spaces. By introducing the S2D-strategy, low-dimensional constellations are sparsely mapped into high-dimensional spaces, effectively reducing the interference caused by superimposed coding. Theoretical analysis shows that sparse dimension allocation can reduce the mutual influence of user signals during decoding, thus improving the system capacity. Simulation experiments further verify the superiority of PD-SDCMA: under the AWGN channel, compared with PD-NOMA, PD-SDCMA can support more user access under QPSK and 16QAM modulations and maintain a lower BER. Future research will combine dynamic channel environments and multi-dimensional modulation technologies such as regular tetrahedron constellation to further optimize the design of the S2D-strategy to adapt to more complex 6G application scenarios.

\vspace{12pt}


\begin{thebibliography}{00}
\bibitem{b1}Y. Akaiwa and H. Andoh, “Channel segregation-a self-organized dynamic channel allocation method: application to TDMA/FDMA microcellular system,” \textit{IEEE Journal on Selected Areas in Communications}, vol. 11, no. 6, pp. 949–954, Jan. 1993.
\bibitem{b2}J. Capetanakis, “Generalized TDMA: The Multi-Accessing Tree Protocol,” \textit{IEEE Transactions on Communications}, vol. 27, no. 10, pp. 1476–1484, Oct. 1979.
\bibitem{b3}W. C. Y. Lee, “Overview of cellular CDMA,” \textit{IEEE Transactions on Vehicular Technology}, vol. 40, no. 2, pp. 291–302, May 1991.
\bibitem{b4}B. Da and C. C. Ko, “Dynamic resource allocation in relay-assisted OFDMA cellular system,” \textit{Transactions on Emerging Telecommunications Technologies}, vol. 23, no. 1, pp. 96–103, Oct. 2011.
\bibitem{b5}Y. Saito, Y. Kishiyama, A. Benjebbour, T. Nakamura, A. Li, and K. Higuchi, “Non-Orthogonal Multiple Access (NOMA) for Cellular Future Radio Access,” \textit{2013 IEEE 77th Vehicular Technology Conference (VTC Spring)}, Jun. 2013.
\bibitem{b6}P. Patel and J. Holtzman, “Analysis of a simple successive interference cancellation scheme in a DS/CDMA system,” \textit{IEEE Journal on Selected Areas in Communications}, vol. 12, no. 5, pp. 796–807, Jun. 1994.
\bibitem{b7}H. Nikopour and H. Baligh, “Sparse code multiple access,” \textit{2013 IEEE 24th Annual International Symposium on Personal, Indoor, and Mobile Radio Communications (PIMRC)}, Sep. 2013.
\bibitem{b8}S. Chen, B. Ren, Q. Gao, S. Kang, S. Sun, and K. Niu, “Pattern Division Multiple Access—A Novel Nonorthogonal Multiple Access for Fifth-Generation Radio Networks,” \textit{IEEE Transactions on Vehicular Technology}, vol. 66, no. 4, pp. 3185–3196, Apr. 2017.
\bibitem{b9}Z. Yuan, G. Yu, W. Li, Y. Yuan, X. Wang, and J. Xu, “Multi-User Shared Access for Internet of Things,” \textit{2016 IEEE 83rd Vehicular Technology Conference (VTC Spring)}, May 2016.
\bibitem{b10} S. Dang, O. Amin, B. Shihada, and M.-S. Alouini, “What should 6G be?,” \textit{Nature Electronics}, vol. 3, no. 1, pp. 20–29, Jan. 2020.
\bibitem{b11} S. Liesegang and S. Buzzi, “Coexistence of eMBB+ and mMTC+ in Uplink Cell-Free Massive MIMO Networks,” 2024, \textit{arXiv:2404.19117v1}.
\bibitem{b12} K. Yu, Z. Feng, J. Yu, T. Chen, J. Peng, and D. Li, “Secure Ultra-Reliable and Low Latency Communication in UAV-Enabled NOMA Wireless Networks,” \textit{IEEE Transactions on Vehicular Technology}, vol. 73, no. 10, pp. 14908–14922, Oct. 2024.
\bibitem{b13} W. Yang \textit{et al}., “Semantic Communications for Future Internet: Fundamentals, Applications, and Challenges,” \textit{IEEE Communications Surveys \& Tutorials}, vol. 25, no. 1, pp. 213–250, 2023.
\bibitem{b14} K. Trichias, A. Kaloxylos, and C. Willcock, “6G Global Landscape: A Comparative Analysis of 6G Targets and Technological Trends,” \textit{2024 Joint European Conference on Networks and Communications \& amp; 6G Summit (EuCNC/6G Summit)}, pp. 1–6, Jun. 2024.
\bibitem{b15} Z. Chen, Z. Zhang, and Z. Yang, “Big AI Models for 6G Wireless Networks: Opportunities, Challenges, and Research Directions,” \textit{IEEE Wireless Communications}, vol. 31, no. 5, pp. 164–172, Oct. 2024.
\bibitem{b16}M. Zeng, A. Yadav, O. A. Dobre, G. I. Tsiropoulos, and H. V. Poor, “Capacity Comparison Between MIMO-NOMA and MIMO-OMA With Multiple Users in a Cluster,” \textit{IEEE Journal on Selected Areas in Communications}, vol. 35, no. 10, pp. 2413–2424, Oct. 2017.
\bibitem{b17}R. Liu  \textit{et al}., “RIS-Empowered Satellite-Aerial-Terrestrial Networks With PD-NOMA,”  \textit{IEEE Communications Surveys \& Tutorials}, vol. 26, no. 4, pp. 2258–2289, Fourthquarter 2024.
\bibitem{b18}X. Mu and Y. Liu, “Exploiting Semantic Communication for Non-Orthogonal Multiple Access,” \textit{IEEE Journal on Selected Areas in Communications}, vol. 41, no. 8, pp. 2563–2576, Aug. 2023.
\end{thebibliography}
\end{document}